\documentclass[aps,prb,twocolumn]{revtex4} 
\usepackage{amsmath} 
\usepackage{graphicx} 
\usepackage{color}

\begin{document} 
\title{Phenomenology of Current-Induced Spin-Orbit Torques} 

\author{Kjetil M.D. Hals and Arne Brataas} 
\affiliation{Department of Physics, Norwegian University of Science and  
Technology, NO-7491, Trondheim, Norway} 
%%%%%%%%%%%%%%%%%%%%%%%%%%%%%%%%%%%%%%%%%%%%%%%%%%%%%%%%%%%%%%%%%%%%%%%%%%%%%%% 
\begin{abstract}
Currents induce magnetization torques via spin-transfer when the spin angular momentum is conserved or via relativistic spin-orbit coupling.  Beyond simple models, the relationship between material properties and spin-orbit torques is not known.
Here, we present a novel phenomenology of current-induced torques that is valid for any strength of intrinsic spin-orbit coupling. In  $\rm Pt|Co|AlO_x$, we demonstrate that the domain walls move in response to
a novel relativistic dissipative torque that is dependent on the domain wall structure and that can be controlled via the Dzyaloshinskii-Moriya interaction. 
Unlike the non-relativistic spin-transfer torque, the new torque can, together with the spin-Hall effect in the Pt-layer, move domain walls by means of electric currents parallel to the walls.
\end{abstract}

\maketitle 

%%%%%%%%%%%%%%%%%%%%%%%%%%%%%%%%%%%%%%%%%%%%%%%%%%%%%%%%%%%%%%%%%%%%%%%%%%%%%%% 
\section{Introduction} 
%%%%%%%%%%%%%%%%%%%%%%%%%%%%%%%%%%%%%%%%%%%%%%%%%%%%%%%%%%%%%%%%%%%%%%%%%%%%%%% 
Electric currents can be used to manipulate the magnetization in ferromagnets.  This phenomenon arises from the intricate coupling between the quasi-particle flow and the collective magnetic degrees of freedom, and it has enabled the development of spin-torque oscillators, spin-logic devices, and spin-transfer torque (STT)-RAM;~\cite{Brataas:nm2012} the latter entered the marketplace at the end of 2012. In its simplest manifestation, 
current-driven excitations of the ferromagnets are caused by the STT mechanism, in which spin angular momentum is transferred from the spin-polarized conduction electrons to the magnetization. Recent theory~\cite{Manchon:prb2008, Garate:prb2009, Hals:epl2010, Pesin:prb2012, Bijl:prb2012, Wang:prl2012} and experiments~\cite{Chernyshov:np2009, Miron:nm2010, Miron:n2011, Miron:nm2011, Garello:arxiv2013, Kurebayashi:arxiv2013} have shown that currents can also cause magnetization torques 
via relativistic, intrinsic  spin-orbit coupling (SOC), often referred to as spin-orbit torques (SOTs). Relativistic SOTs provide alternative and efficient routes for the manipulation of the magnetization that, in contrast to 
non-relativistic STTs, require neither spin-polarizers nor textured ferromagnets. 

SOTs exist in systems with structural asymmetry, which induces a strong intrinsic SOC.
Theory predicts that SOC in combination with an applied electric field  produces an out-of-equilibrium spin density that in turn yields a torque on the magnetization.~\cite{Manchon:prb2008, Chernyshov:np2009, Garate:prb2009, Hals:epl2010, Pesin:prb2012, Bijl:prb2012, Wang:prl2012} The effect of an SOT was first  detected in strained (Ga,Mn)As.~\cite{Chernyshov:np2009}
The observed switching of samples with uniform magnetization markedly differs from the typical behavior
in metallic ferromagnets, wherein a heterogeneous magnetic texture is required to induce an STT.~\cite{Brataas:nm2012,Ralph:jmmm2008} A similar switching of a single-domain ferromagnet has also been observed in an ultrathin Co layer sandwiched between Pt and $\rm AlO_x$.~\cite{Miron:nm2010, Miron:n2011, Liu:prl2012} Both SOTs~\cite{Manchon:prb2008, Miron:nm2010, Miron:n2011} and the spin-Hall effect (SHE) in combination with conventional STTs~\cite{Liu:prl2012, Haazen:nm2013, Emori:arxiv2013} have been proposed to be responsible for the switching Co magnetization. In the latter case, the SHE in the Pt layer produces a spin-current that diffuses into the Co layer, causing an STT on the magnetization.~\cite{Liu:prl2012}  Interestingly, in this system, experiments have observed fast current-driven motion of ferromagnetic domain walls (DWs).~\cite{Moore:apl2008, Miron:nm2011, Emori:arxiv2013} An SOC-induced antisymmetric exchange interaction, known as the Dzyaloshinskii-Moriya interaction (DMI),~\cite{Dzyaloshinsky:jpcs1958, Moriya:pr1960} stabilizes chiral DWs~\cite{Heide:prb2008, Thiaville:epl2012} with a high DW mobility whose spin texture is very robust with respect to high current densities.~\cite{Moore:apl2008, Miron:nm2011, Thiaville:epl2012, Emori:arxiv2013, Kim:prb2012, Seo:apl2012, Khvalkovskiy:prb2013} Highly efficient, current-driven DW motion has also been observed in trilayers of Co and Ni interfaced with Pt.~\cite{Ryu:ape2012, Thomas:nnt2013} In both $\rm Pt|Co|AlO_x$ and $\rm Co|Ni|Co$ systems, the DWs move at high speeds and occasionally in a direction opposite of that expected for the non-relativistic bulk STT. 

The aforementioned experiments demonstrate that ultrathin metallic ferromagnets in asymmetric heterostructures are important in the design of future spintronic devices wherein the magnetization is efficiently controlled by electric currents. The interplay between SOC effects, such as the DMI, SOTs, and SHE, is believed to be responsible for the remarkable behavior of these systems. However, a detailed understanding of their properties requires improved theoretical models that exceed the present phenomenological framework used to model current-induced magnetization dynamics.

In this Letter, we formulate a general phenomenology of current-induced torques in systems with arbitrarily strong spin-orbit interaction. We apply the formalism to compute the current-driven DW drift velocity in $\rm Pt|Co|AlO_x$ trilayers and to identify a novel SOT. The effect of the new torque can be manipulated via the DMI and can, together with the SHE, induce DW motion even when the electric currents are applied parallel to the DWs. This opens a completely new path for controlling the dissipative current-induced torque by engineering the interfaces in ferromagnetic heterostructures.

The magnetization dynamics of metallic ferromagnets is  well-described by the 
Landau-Lifshitz-Gilbert-Slonczewski (LLGS) equation:~\cite{Ralph:jmmm2008}
\begin{equation}
\dot{\mathbf{m}} = -\gamma \mathbf{m}  \times \mathbf{H}_{\rm eff} + \mathbf{m}  \times \boldsymbol{\alpha} \dot{ \mathbf{m}}  + \boldsymbol{\tau} . \label{Eq:LLG}
\end{equation}
Here, $\gamma$ is (minus) the gyromagnetic ratio, $\mathbf{m} (\mathbf{r}, t)= \mathbf{M}/M_s$ ($M_s= |\mathbf{M}|$) is the unit direction vector of the magnetization $\mathbf{M} (\mathbf{r}, t)$, and $\mathbf{H}_{\rm eff}= - \delta F[\mathbf{M}]/\delta \mathbf{M}$ is the effective field determined by the magnetic free-energy functional $ F[\mathbf{M}]$. The friction processes in the magnetic system are modeled by the Gilbert damping that is parameterized by the symmetrical, second-rank tensor $\boldsymbol{\alpha}$. The last term on the right-hand side of Eq.~\eqref{Eq:LLG} is 
~\cite{Ralph:jmmm2008} 
\begin{equation}
\boldsymbol{\tau} (\mathbf{r}, t)  =  - \left( 1 - \beta\mathbf{m}\times   \right)  \left( \mathbf{v}_s\cdot\boldsymbol{\nabla} \right) \mathbf{m}, \label{Eq:STT}
\end{equation}
and describes the current-induced torques. The torques of Eq.~\eqref{Eq:STT} contain reactive and dissipative contributions. The term 
proportional to the parameter $\beta$ represents the dissipative torque because it breaks 
the time-reversal symmetry of Eq.~\eqref{Eq:LLG}. The vector $\mathbf{v}_s$ is directed along the out-of-equilibrium current density $\boldsymbol{\mathcal{J}}$ and is proportional to  
the current density, the equilibrium spin density,  and the spin polarization of the induced current. Eq.~\eqref{Eq:STT} is fully rotationally symmetric in both  the
spin space and the coordinate space. This can be seen by applying a global rotation operator $\mathcal{R}$ to the magnetization and the spatial vectors: $\tilde{\mathbf{m}}= \mathcal{R}\mathbf{m}$, $\tilde{\mathbf{v}}_s= \mathcal{R}\mathbf{v}_s$, and $\tilde{\boldsymbol{\nabla}}= \mathcal{R} \boldsymbol{\nabla}$. Separate rotations of the spin space and the coordinate space leave the magnitude of the torque invariant:
\begin{eqnarray}
| \left( 1 - \beta\tilde{\mathbf{m}}\times   \right)  \left( \mathbf{v}_s\cdot\boldsymbol{\nabla} \right) \tilde{\mathbf{m}} | 
&=& |  \left( 1 - \beta\mathbf{m}\times   \right)  \left( \tilde{\mathbf{v}}_s\cdot\tilde{\boldsymbol{\nabla}} \right) \mathbf{m} | \nonumber .
\end{eqnarray}
Thus, the STT given by Eq.~\eqref{Eq:STT} is completely decoupled from the symmetry of the underlying crystal lattice. 

However, in systems with strong intrinsic SOC, Eq.~\eqref{Eq:STT} is not sufficient to describe the magnetization dynamics. 
The only possible effect of the intrinsic SOC included in Eq.~\eqref{Eq:STT}  is the renormalization of the $\beta$ parameter.
The intrinsic SOC mediates a coupling between the spin space and the crystal lattice. As a consequence, the system is only invariant under proper and improper rotations that leave the crystal structure unchanged and that act simultaneously on the spin space and on the coordinate space.
The intrinsic SOC therefore introduces several new torques that are dependent on the orientation of the magnetization with respect to the crystal lattice. Examples of such SOTs have been derived theoretically in Refs.~\onlinecite{Manchon:prb2008, Garate:prb2009, Hals:epl2010,Bijl:prb2012,Garello:arxiv2013}.  However, to date, a general theoretical framework has not been developed that systematically and correctly extends the LLG phenomenology in Eq.~\eqref{Eq:LLG} to include the intrinsic SOC effects.

%%%%%%%%%%%%%%%%%%%%%%%%%%%%%%%%%%%%%%%%%%%%%%%%%%%%%%%%%%%%%%%%%%%%%%%%%%%%%%%%%%%%%%%%%%%%%%%%%%%%%%%%%%%%%%%% 
\section{Theory} 
%%%%%%%%%%%%%%%%%%%%%%%%%%%%%%%%%%%%%%%%%%%%%%%%%%%%%%%%%%%%%%%%%%%%%%%%%%%%%%%%%%%%%%%%%%%%%%%%%%%%%%%%%%%%%%% 
To derive our theory, we consider the most general form of  the current-induced torque in the local approximation: 
\begin{equation}
\boldsymbol{\tau} (\mathbf{r}, t) = \mathbf{m}\times \mathbf{H}_{\rm c} [\mathbf{m}, \boldsymbol{\nabla} \mathbf{m}, \boldsymbol{\mathcal{J}} ], \label{Eq:SOT_0}
\end{equation}
where $\mathbf{H}_{\rm c}$ is an effective field induced by the out-of-equilibrium  current density that is dependent on the local magnetization direction and the local gradients of the magnetization. 
Here, $\boldsymbol{\nabla} \mathbf{m}$ is a shorthand notation for all the different combinations of $\partial_j m_i $ ($\partial_j m_i \equiv \partial m_i / \partial r_j $). In the absence of an out-of-equilibrium current density, $\mathbf{H}_c$ vanishes.  
Because our interest is the linear response regime, we 
expand Eq.~\eqref{Eq:SOT_0} to the first order in $\boldsymbol{\mathcal{J}} $  
\begin{equation}
\boldsymbol{\tau} (\mathbf{r}, t) = \mathbf{m}\times \boldsymbol{\eta}\, \boldsymbol{\mathcal{J}}. \label{Eq:SOT_1}
\end{equation}
Here, the second-rank tensor $\left( \boldsymbol{\eta} [\mathbf{m}, \boldsymbol{\nabla} \mathbf{m}]\right)_{ij} = \left( \partial H_{c,i}/ \partial \mathcal{J}_j \right)_{\boldsymbol{\mathcal{J}}=0}$ operates on the vector $\boldsymbol{\mathcal{J}}$. $\boldsymbol{\eta} (\mathbf{r}, t)$ completely determines the symmetry of the local torque $\boldsymbol{\tau} (\mathbf{r}, t)$ and includes all of the current-induced torque effects.  Our main aim is to derive a general and simplified expression for $\boldsymbol{\eta} (\mathbf{r}, t)$.

\begin{figure}[ht] 
\centering 
\includegraphics[scale=1.0]{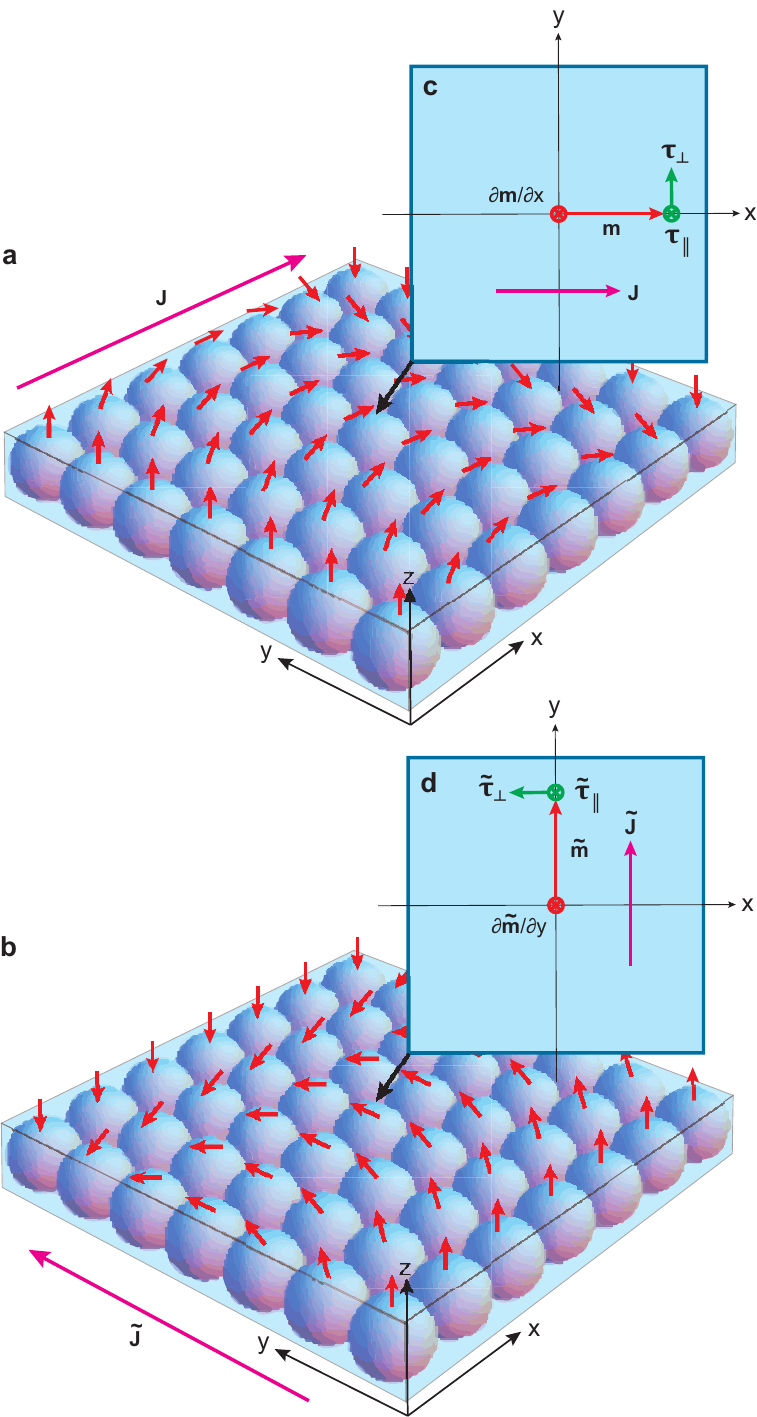}  
\caption{(color online). The magnetic textures and current densities in (a) and (b) are related by a rotation of $90$ degrees about the $z$ axis.
Due to the four-fold symmetry of the underlying crystal lattice, the two cases are equivalent. Neumann's principle implies that the local torques on the two textures are related by a four-fold rotation. This is illustrated in (c)  and 
(d), which show the torque, the gradient, and the direction of the magnetization at two equivalent points indicated by the black arrows.   }
\label{Fig1} 
\end{figure} 

We will deduce the second-rank tensor $\boldsymbol{\eta}$ from Neumann's principle, which states that "\emph{any type of symmetry which is exhibited by the point group of the crystal is possessed by every physical property of the crystal}".~\cite{Birss:book} An illustration of how this principle imposes symmetry relations on the torque in Eq.~\eqref{Eq:SOT_1} is presented Fig.~\ref{Fig1}. As an example, let us assume that the
material has a four-fold symmetry axis along $z$. Figs.~\ref{Fig1}a - b show two systems containing a DW along the $y$ and $x$ axis, respectively, with a current density applied perpendicular to the DW. The two systems in Figs.~\ref{Fig1}a - b 
are related by a rotation $\mathcal{R}_{4z}$  of $90$ degrees about the $z$ axis,  
which leads to the following transformations: $\tilde{\mathbf{r}}= \mathcal{R}_{4z}\mathbf{r}$, 
 $\tilde{\boldsymbol{\nabla}}= \mathcal{R}_{4z}\boldsymbol{\nabla}$, $\tilde{\boldsymbol{\mathcal{J}}}= \mathcal{R}_{4z}\boldsymbol{\mathcal{J}}$, and $\tilde{\mathbf{m}}= |\mathcal{R}_{4z}|\mathcal{R}_{4z}\mathbf{m}$.
Because the magnetization is a pseudovector, its transformation rule includes the determinant $|\mathcal{R}_{4z}|$.   
Due to the  four-fold symmetry of the system, the two cases in Figs.~\ref{Fig1}a - b are equivalent, and Neumann's principle implies that the torques induced on the two magnetic textures are 
related by $\mathcal{R}_{4z}$, {\it i.e.}, $\tilde{\boldsymbol{\tau}}= |\mathcal{R}_{4z}| \mathcal{R}_{4z} \boldsymbol{\tau}$ (Figs.~\ref{Fig1}c - d). Thus, by applying Eq.~\eqref{Eq:SOT_1}, we find that 
 $\tilde{\mathbf{m}}\times  \boldsymbol{\eta} [\tilde{\mathbf{m}}, \tilde{\boldsymbol{\nabla}}\tilde{ \mathbf{m}} ]\tilde{\boldsymbol{\mathcal{J}}}  = |\mathcal{R}_{4z}| \mathcal{R}_{4z} \left( \mathbf{m}\times  \boldsymbol{\eta} [\mathbf{m}, \boldsymbol{\nabla} \mathbf{m} ]  \boldsymbol{\mathcal{J}} \right) $,  which yields the following symmetry relations for $\boldsymbol{\eta}$:   
 \begin{equation}
\boldsymbol{\eta}  [\tilde{\mathbf{m}}, \tilde{\boldsymbol{\nabla}}\tilde{ \mathbf{m}} ]  = |\mathcal{R}| \mathcal{R} \boldsymbol{\eta} [\mathbf{m}, \boldsymbol{\nabla} \mathbf{m} ]  \mathcal{R}^T . \label{Eq:SOT_2}
\end{equation}
This is the first central result of our work. We have removed the subscript of $ \mathcal{R}$ denoting the symmetry operation because Eq.~\eqref{Eq:SOT_2} holds for any symmetry operation that is an element of the crystal's point group. Although Eq.~\eqref{Eq:SOT_2} determines many symmetry aspects of the torque, the dependence of torque on the magnetization direction and the gradients of the magnetization means that a large number of functions can satisfy Eq.~\eqref{Eq:SOT_2}. This situation closely resembles the case of the large number of terms that can describe the magnetocrystalline anisotropy energy of a system. Although the anisotropy energy should obey the symmetries of the material, it is common to approximate the dependence by the first harmonics of this dependence. 

Thus, to simplify the tensor form of $\boldsymbol{\eta}$, we assume that its components can be expressed as a series expansion in increasing 
powers of $m_i$ and $\partial_j m_i$ 
\begin{equation}
\eta_{ij} = \Lambda^{(r)}_{ij} + \Lambda^{(d)}_{ijk} m_k + \beta_{ijkl}\partial_k m_l + P_{ijkln}m_k \partial_l m_n  + ...  \label{Eq:eta}
\end{equation}
In Eq.~\eqref{Eq:eta}, and in the following discussion, summation over repeated indices is implied.
The power expansion in Eq.~\eqref{Eq:eta} represents a systematic scheme for deriving current-induced torques of increasing degrees of anisotropy.
The number of terms  in the expansion required to obtain a sufficiently accurate  
description of the magnetization dynamics can be determined from experimental data.  As a demonstration, 
we limit our discussion to  the terms explicitly written in Eq.~\eqref{Eq:eta} because they lead to a simple extension of Eq.~\eqref{Eq:LLG}, and similar to the case of magnetic anisotropy, we expect the first harmonics to provide a sufficient description of many materials.

In Eq.~\eqref{Eq:eta}, $\boldsymbol{\Lambda^{(r)}}$ and $\boldsymbol{\Lambda^{(d)}}$ describe the reactive and dissipative SOTs, respectively, that are present even in systems with uniform magnetization. 
The tensors $\mathbf{P}$ and $\boldsymbol{\beta}$  are generalizations of the reactive and dissipative torques given in Eq.~\eqref{Eq:STT}. 
Our theory therefore demonstates that in systems with strong intrinsic SOC, the reactive torque is generalized to become 
a tensor of rank five, while the dissipative torque parameter $\beta$ becomes a tensor of rank four.  From our derived symmetry relations in Eq.~\eqref{Eq:SOT_2} and the expansion of $\boldsymbol{\eta} $ in Eq.~\eqref{Eq:eta}, we find that the tensors  $\boldsymbol{\Lambda^{(r)}}$, $\boldsymbol{\Lambda^{(d)}}$, $\boldsymbol{\beta}$, and $\mathbf{P}$ must satisfy the symmetry relations   
\begin{eqnarray}
\Lambda^{(r)}_{ij} &=& |\mathcal{R}| \mathcal{R}_{i i^{'}} \mathcal{R}_{j j^{'}}  \Lambda^{(r)}_{i^{'} j^{'}} , \label{Eq:SymRel_1} \\
\Lambda^{(d)}_{ijk} &=&  \mathcal{R}_{i i^{'}} \mathcal{R}_{j j^{'}} \mathcal{R}_{k k^{'}}  \Lambda^{(d)}_{i^{'} j^{'} k^{'}} , \label{Eq:SymRel_2} \\
\beta_{ijkl} &=&  \mathcal{R}_{i i^{'}} \mathcal{R}_{j j^{'}} \mathcal{R}_{k k^{'}} \mathcal{R}_{l l^{'}}  \beta_{i^{'} j^{'} k^{'} l^{'}}, \label{Eq:SymRel_3} \\
P_{ijkln} &=& |\mathcal{R}|  \mathcal{R}_{i i^{'}} \mathcal{R}_{j j^{'}} \mathcal{R}_{k k^{'}} \mathcal{R}_{l l^{'}} \mathcal{R}_{n n^{'}}  P_{i^{'} j^{'} k^{'} l^{'} n^{'}}. \label{Eq:SymRel_4}
\end{eqnarray}
Eqs.~\eqref{Eq:SymRel_1} - \eqref{Eq:SymRel_4} represent our second central result, and they reduce the number of independent tensor coefficients and greatly simplify the forms of the tensors.
For example, in systems that are invariant under the inversion operator $\mathcal{R}_{ij}= -\delta_{ij}$, Eqs.~\eqref{Eq:SymRel_1} - \eqref{Eq:SymRel_2} imply that $\boldsymbol{\Lambda^{(r)}}= \boldsymbol{\Lambda^{(d)}}= 0$.
This means that SOTs are absent in inversion-symmetric systems with uniform magnetization.~\cite{Manchon:prb2008, Chernyshov:np2009,Garate:prb2009, Hals:epl2010}

The tensors determined from Eqs.~\eqref{Eq:SOT_2} - \eqref{Eq:eta}, together with Eq.~\eqref{Eq:SOT_1}, represent a general phenomenology of current-induced magnetization dynamics that correctly accounts for 
intrinsic SOC effects. The torques originating from the low-order tensors given by Eqs.~\eqref{Eq:SymRel_1} - \eqref{Eq:SymRel_4} lead to a generalization of Eq.~\eqref{Eq:STT}.
The torques in Eq.~\eqref{Eq:STT} are obtained from our formalism by assuming a full rotational symmetry under separate rotations of the spin space and the coordinate space.  
Under these symmetry requirements, Eqs.~\eqref{Eq:SymRel_1} - \eqref{Eq:SymRel_4}  imply that  $\boldsymbol{\Lambda^{(r)}}= \boldsymbol{\Lambda^{(d)}}= 0$, $P_{ijkln}= P \delta_{jl} \epsilon_{ikn}$, and $\beta_{ijkl}= P \beta \delta_{jk} \delta_{il}$, which 
lead to the non-relativistic STT in Eq.~\eqref{Eq:STT} when $P\boldsymbol{\mathcal{J}}= \mathbf{v}_s$. Here, $\delta_{jn}$ and $\epsilon_{ilk}$ are the Kronecker delta and the Levi-Civita tensor, respectively. 

%%%%%%%%%%%%%%%%%%%%%%%%%%%%%%%%%%%%%%%%%%%%%%%%%%%%%%%%%%%%%%%%%%%%%%%%%%%%%%% 
\section{Application: Ferromagnetic Heterostructure}
%%%%%%%%%%%%%%%%%%%%%%%%%%%%%%%%%%%%%%%%%%%%%%%%%%%%%%%%%%%%%%%%%%%%%%%%%%%%%%% 
Next, we consider the ferromagnetic heterostructure $\rm Pt|Co|AlO_x$ and apply the phenomenology to the study of current-induced DW motion. The growth of the ultrathin Co layer, {\it i.e.}, the stacking direction, occurs along the crystallographic direction $[111]$.~\cite{Daalderop:prb1994} We assume a fully epitaxial system. The single Co layer has a trigonal crystal structure that is described by the centrosymmetric point group $\rm D_{3d}$, in which the three-fold axis of rotation is along the $[111]$ direction.~\cite{Daalderop:prb1994} The Pt and $\rm AlO_x$ layers break the inversion symmetry of the system and reduce the symmetry group to $\rm C_{3v}$. The SOC in the Co layer is, therefore, assumed to be invariant under the action of $\rm C_{3v}$ when the symmetry operations act simultaneously on the spin and coordinate spaces. Thus, the allowed torques on the magnetization in the Co layer are determined from Eq.~\eqref{Eq:SOT_2} with $\mathcal{R}\in \rm C_{3v}$. In the following description, we denote the $[111]$ direction as the $z$ axis, and the $x$ axis is defined such that the $xz$ plane corresponds to one of the three equivalent reflection planes of $\rm C_{3v}$.  

The magnetic Co layer is modeled by the free energy density~\cite{Thiaville:epl2012}
\begin{eqnarray}
\mathcal{F} &=&  (J/2) \partial_i \mathbf{m}\cdot \partial_i \mathbf{m}  + D\left( m_z \partial_i m_i - m_i \partial_i m_z \right) + \nonumber  \\
& &  (K_{\nu}/2) \left(\mathbf{m}\cdot\boldsymbol{\nu}  \right)^2 -  (K_{z}/2) m_z^2.
\end{eqnarray}
Here, $i\in \left\{ x, y \right\}$, $J$ is the spin stiffness, $K_z > 0$ and $K_{\nu} > 0$ are anisotropy constants,  and the term proportional to $D$ is the DMI. 
$\boldsymbol{\nu}= \left[ \cos (\phi_0), \sin (\phi_0), 0 \right]$ describes the direction of the DW.
To model the DW dynamics, we apply a collective coordinate description in which the DW is determined 
by the ansatz $\mathbf{m} = \left[ {\rm sech} (\mu) \cos (\phi_m), {\rm sech} (\mu) \sin (\phi_m), \tanh (\mu)   \right]$, where $\mu = (\nu - r_w)/\lambda_w$. Here, $\lambda_w$ is the DW length, which is assumed to be static, 
and $r_w=r_w(t)$ and $\phi_m= \phi_m (t)$ are the collective coordinates that describe the DW position and the tilting angle of the texture, respectively. 
In the absence of any external fields,  the equilibrium value of $\phi_m$ is given by 
\begin{equation}
\cos ( \phi_m^{\rm eqv} - \phi_0 ) = \pi D / 2K_{\nu}\lambda_w. 
\end{equation}
If $\pi D / 2K_{\nu}\lambda_w \geq 1$ or $\pi D / 2K_{\nu}\lambda_w \leq -1$, the equilibrium angle is $\phi_m^{\rm eqv} = \phi_0$ or $\phi_m^{\rm eqv} = \phi_0 + \pi$, respectively.

We believe that the DW dynamics are dominated by 
the lowest-order terms given by Eqs.~\eqref{Eq:SymRel_1} - \eqref{Eq:SymRel_4} in combination with the SHE torque;
higher-order terms can be included to refine the results. 
In contrast, a recent experiment on a homogenous $\rm Pt|Co|AlO_x$ system indicates that higher-order terms in the series expansion of $\boldsymbol{\eta}$ in Eq.~\eqref{Eq:eta}  may influence the dynamics of a single magnetic domain.~\cite{Garello:arxiv2013}  The SHE torque is modeled by the Slonczewski torque  $\boldsymbol{\tau}_{\rm she}= \tau_{\rm she}^{(0)} \mathbf{m} \times (\mathbf{s}\times\mathbf{m})$, where $\mathbf{s}= \boldsymbol{\mathcal{J}}\times \hat{\mathbf{z}}$ is 
proportional to the polarization of the spin current induced within the Pt layer by the electric current density $\boldsymbol{\mathcal{J}}= \mathcal{J}_0 [ \cos (\phi_e),  \sin (\phi_e), 0]$.~\cite{Liu:prl2012}  
For the symmetry group $\rm C_{3v}$, the tensors $\boldsymbol{\Lambda^{(r)}}$, $\boldsymbol{\Lambda^{(d)}}$, and $\boldsymbol{\beta}$ are described by one, five, and 14 independent tensor coefficients, respectively, and
their explicit forms can be found in Ref.~\onlinecite{Birss:book}.  In Eq.~\eqref{Eq:SymRel_4}, we keep only the fully isotropic part $P_{ijkln} \sim  \delta_{jl} \epsilon_{ikn}$. Its value is largely determined by the spin polarization of the induced current and the equilibrium spin density. The remaining part of $P_{ijkln}$ is governed by the SOC and is therefore assumed to provide a negligible correction. A similar simplification cannot be applied to the $\beta_{ijkl}$ tensor because even the isotropic part   
$\beta_{ijkl}\sim \delta_{jk} \delta_{il}$ is proportional to the spin-flip rate, which is primarily controlled by the SOC. The Gilbert damping tensor is diagonal and determined by the two parameters $\alpha_{xx}=\alpha_{yy}= \alpha_{\bot}$ and $\alpha_{zz}$. 

The equations of motion for $r_w$ and $\phi_m$ are~\cite{Tretiakov:prl2008}  
\begin{equation}
 \left( \Gamma_{ij} - G_{ij} \right) \dot{a}_j = \gamma F_i +  \sum_{a=1}^5 L_i^{(a)}, \label{Eq:CC} 
\end{equation}
where $a_i \in \left\{ r_w, \phi_m \right\}$.  
The matrices $\Gamma_{ij}$ and $G_{ij}$ are given by $G_{ij} = \int  \mathbf{m}\cdot [  (\partial \mathbf{m} / \partial a_i)\times (\partial \mathbf{m} / \partial a_j) ] {\rm dV}$ and
$\Gamma_{ij} = \int  (\partial \mathbf{m} / \partial a_i)\cdot \boldsymbol{\alpha} (\partial \mathbf{m} / \partial a_j)  {\rm dV}$, and $F_i$ is the force due to the effective field, 
$F_i = \int  \mathbf{H}_{\rm eff}\cdot (\partial \mathbf{m} / \partial a_i)  {\rm dV}$. The current-induced dynamics is described by the five terms 
$L_i^{(a)} =    \int  (\partial \mathbf{m} / \partial a_i) \cdot [  \mathbf{m} \times \boldsymbol{\tau}^{(a)} ] {\rm dV} $, where the superscript $(a)$ denotes the four torques determined from Eqs.~\eqref{Eq:SymRel_1} - \eqref{Eq:SymRel_4} and the SHE torque. Solving Eq.~\eqref{Eq:CC} in the linear response regime, we compute our final central result, the steady-state DW velocity
\begin{eqnarray}
\dot{r}_w &=& \frac{1}{2}\frac{\mathcal{J}_0}{\alpha_{\bot} + 2 \alpha_{zz}} [  \tau_{\rm soc}\cos (\phi_e - \phi_m^{\rm eqv}) \label{Eq:drdt}  \\
& & +  \beta_{\rm iso} \cos (\phi_e - \phi_0) +  \beta_{m} \cos (\phi_e + \phi_0 - 2\phi_m^{\rm eqv})   ] , \nonumber
\end{eqnarray}
where $\tau_{\rm soc} = (3\pi \lambda_w /2) (  \Lambda^{(d)}_{zyy} - \Lambda^{(d)}_{xxz} + 2 \tau_{\rm she}^{(0)} )$,  $\beta_{\rm iso} = \beta_{xxyy} +\beta_{xyxy}  + 2\beta_{xyyx}  + 4\beta_{zxxz} $, and
$\beta_m= \beta_{xxyy} + \beta_{xyxy}$. The first term in Eq.~\eqref{Eq:drdt} arises from the SHE torque and the dissipative homogeneous SOT. The term proportional to 
$\beta_{\rm iso}$ represents a conventional dissipative STT term in which the current-driven DW motion depends only on the relative angle between the applied current and the DW direction. 

The expression for $\dot{r}_w$ of Eq.~\eqref{Eq:drdt} also contains a new dissipative current-induced torque that has not been previously reported. The term proportional to
$\beta_m$ describes a dissipative torque that is dependent on the DW structure through the angle $\phi_m^{\rm eqv}$. 
This is a purely relativistic torque because its magnitude depends on the tensor coefficients $\beta_{xxyy}$ and $\beta_{xyxy}$, which vanish in the absence of intrinsic SOC.
The angle $\phi_m^{\rm eqv}$ is dependent on the DMI and can be manipulated by engineering the $\rm Pt|Co$ and $\rm Co|AlO_x$ interfaces, as demonstrated experimentally in Ref.~\onlinecite{Thomas:nnt2013}. The $\beta_m$ term therefore opens an exciting new path for controlling the effective dissipative torque on the DW via the DMI. In principle, even the sign of the total dissipative torque arising from the $\beta_{ijkl}$ tensor can be controlled by the DMI if $\beta_m$ is of the same order of magnitude as $\beta_{\rm iso}$. 

Another interesting phenomenon observed from Eq.~\eqref{Eq:drdt} is the current-driven DW motion induced by an electric current applied perpendicular to the texture direction, {\it i.e.}, $\boldsymbol{\mathcal{J}}\,\bot\,\boldsymbol{\nu}$. Depending on the value of $\phi_m^{\rm eqv}$, both the $\tau_{\rm soc}$ term and the $\beta_m$ term can induce a DW motion in this case. However, the effect requires that $|\pi D / 2K_{\nu}\lambda_w| < 1$. 
When $\boldsymbol{\mathcal{J}}\,\bot\,\boldsymbol{\nu}$, the torque efficiency of the $\tau_{\rm soc}$ term is maximized for $\phi_m^{\rm eqv}= \phi_e~({\rm mod~\pi})$, {\it i.e.}, a Bloch wall structure, whereas the $\beta_m$ term attains its maximum efficiency for the angles $ \phi_m^{\rm eqv} = \phi_0 + \pi / 4 ~ ({\rm mod}~\pi/2)$. For $\boldsymbol{\mathcal{J}}\, ||  \,\boldsymbol{\nu}$, Eq.~\eqref{Eq:drdt} implies that a Bloch wall motion is only induced by the two $\beta$ terms, whereas both the    
$\tau_{\rm soc}$ term and the $\beta$ terms contribute to the motion of N\'eel walls, {\it i.e.}, when $\phi_m^{\rm eqv} = \phi_0~({\rm mod~\pi})$. This is in agreement with previously reported results.~\cite{Emori:arxiv2013, Thomas:nnt2013}          

In polycrystalline or disordered systems,  the discrete three-fold rotation symmetry is averaged out, and a full rotational symmetry about the $z$ axis can be assumed. However, this 
symmetry still allows for all tensor coefficients in Eq.~\eqref{Eq:drdt} to be present, and the equation remains unchanged. 

%%%%%%Summary%%%%%%%%%%%%%%%%%%%%
\section{Summary}
In summary, we developed a novel phenomenology of current-induced magnetization dynamics that is valid for any strength of the intrinsic SOC.
The formalism is applied to the study of current-induced DW motion in the ferromagnetic heterostructure $\rm Pt|Co|AlO_x$.
Our results show that the current-driven DW motion is dependent on a novel dissipative STT that can be controlled via the DMI. 

%%% References %%%%%%%%%%%%%%%%%%%%%%%%%%%%%%%%%%%%%%%%%%%%%%%%%%%%%%%%%%%% 


\begin{thebibliography}{99} 

\bibitem{Brataas:nm2012} A. Brataas, A. D. Kent, and H. Ohno, Nature Mat. {\bf 11}, 372 (2012); D. C. Ralph and M. Stiles, J. Magn. Mat. {\bf 320}, 1190 (2008). 

\bibitem{Manchon:prb2008} A. Manchon and S. Zhang, Phys. Rev. B {\bf 78}, 212405 (2008).

\bibitem{Garate:prb2009} I. Garate and A. H. MacDonald, Phys. Rev. B {\bf 80}, 134403 (2009). 

\bibitem{Hals:epl2010} K. M. D. Hals, A. Brataas, and Y. Tserkovnyak, Europhys. Lett. {\bf 90}, 47002 (2010).

\bibitem{Pesin:prb2012} D. A. Pesin and A. H. MacDonald, Phys. Rev. B {\bf 86}, 014416 (2012).

\bibitem{Bijl:prb2012} E. van der Bijl and R. A. Duine, Phys. Rev. B {\bf 86}, 094406 (2012). 

\bibitem{Wang:prl2012}X. Wang and A. Manchon, Phys. Rev. Lett. {\bf 108}, 117201 (2012).

\bibitem{Chernyshov:np2009} A. Chernyshov et al., Nat. Phys. {\bf 5}, 656 (2009).

\bibitem{Miron:nm2010} I. M. Miron {\it et al.}, Nat. Mat. {\bf 9}, 230 (2010).

\bibitem{Miron:n2011} I. M. Miron {\it et al.}, Nature (London) {\bf 476}, 189 (2011).

\bibitem{Miron:nm2011} I. M. Miron {\it et al.}, Nat. Mat. {\bf 10}, 419 (2011).

\bibitem{Garello:arxiv2013} K. Garello {\it et al.},  arXiv:1301.3573.

\bibitem{Kurebayashi:arxiv2013} H. Kurebayashi {\it et al.},  arXiv:1306.1893.

\bibitem{Ralph:jmmm2008}For reviews see, e.g., D. C. Ralph and M. Stiles, J. Magn. Magn. Mater. {\bf 320}, 1190 (2008).

\bibitem{Liu:prl2012} L. Liu, O. J. Lee, T. J. Gudmundsen, D. C. Ralph, and R. A. Buhrman, Phys. Rev. Lett. {\bf 109}, 096602 (2012).

\bibitem{Haazen:nm2013} P. P. J. Haazen {\it et al.}, Nat. Mat. {\bf 12},  299 (2013).

\bibitem{Emori:arxiv2013} S. Emori, U. Bauer, S. M Ahn, E. Martinez, and G. S. D. Beach, Nat. Mat. {\bf 12}, 611 (2013).

\bibitem{Moore:apl2008}T. A. Moore {\it et al.}, Appl. Phys. Lett. {\bf 93}, 262504 (2008).

\bibitem{Dzyaloshinsky:jpcs1958} I. Dzyaloshinsky, J. Phys. Chem. Solids {\bf 4}, 241 (1958).

\bibitem{Moriya:pr1960}T. Moriya, Phys. Rev. {\bf 120}, 91 (1960); Phys. Rev. Lett. {\bf 4}, 228 (1960).

\bibitem{Heide:prb2008}M. Heide, G. Bihlmayer, and S. Bl\"ugel, Phys. Rev. B {\bf 78}, 140403 (2008).

\bibitem{Thiaville:epl2012}A. Thiaville, S. Rohart, E. Jue, V. Cros, and A. Fert, Europhys. Lett. {\bf 100}, 57002 (2012).

\bibitem{Kim:prb2012}K. W. Kim, S. M. Seo, J. Ryu, K. J. Lee, and H. W. Lee, Phys. Rev. B {\bf 85}, 180404 (2012).

\bibitem{Seo:apl2012}S. M. Seo, K. W. Kim, J. Ryu, H. W. Lee, and K. J. Lee, Appl. Phys. Lett. {\bf 101}, 022405 (2012).

\bibitem{Khvalkovskiy:prb2013}A. V. Khvalkovskiy {\it et al.}, Phys. Rev. B {\bf 87}, 020402 (2013). 

\bibitem{Ryu:ape2012}K. S. Ryu, L. Thomas, S. H. Yang, and S. S. P. Parkin, Appl. Phys. Expr. {\bf 5}, 093006 (2012).

\bibitem{Thomas:nnt2013} K. S. Ryu, L. Thomas, S. H. Yang, and S. Parkin,  Nat. Nanotech. {\bf 8}, 527 (2013). 

\bibitem{Birss:book}R. R. Birss, \emph{ Symmetry and Magnetism} (North-Holland, Amsterdam, 1966).

\bibitem{Daalderop:prb1994} G. H. O. Daalderop, P. J. Kelly, and M. F. H. Schuurmans, Phys. Rev. B. {\bf 50}, 9989 (1994).

\bibitem{Tretiakov:prl2008} O. A. Tretiakov,  D. Clarke, Gia-Wei Chern, Ya. B. Bazaliy, and O. Tchernyshyov, Phys. Rev. Lett. {\bf 100}, 127204 (2008).

\end{thebibliography}
\end{document}